\documentclass[letterpaper]{article}
\usepackage{mume}
\usepackage{paralist}
\usepackage{times}
\usepackage{helvet}
\usepackage{courier}
\usepackage{graphicx}
\usepackage{booktabs}
\usepackage{microtype}
\usepackage{flushend}
\usepackage{color}

\begin{document}
	% The file mume.sty is the style file for MUME 
	\title{Explicitly Conditioned Melody Generation: \\
    A Case Study with Interdependent RNNs}
	\author{Benjamin Genchel\\
    Center for Music Technology\\
    Georgia Institute of Technology\\
    Atlanta, GA 30332 USA\\ 
    bgenchel3@gatech.edu\\
    \And
    Ashis Pati\\
    Center for Music Technology\\
    Georgia Institute of Technology\\
    Atlanta, GA 30332 USA\\
    ashis.pati@gatech.edu\\
    \And
    Alexander Lerch\\
    Center for Music Technology\\
    Georgia Institute of Technology\\
    Atlanta, GA 30332 USA\\
    alexander.lerch@gatech.edu\\
    }
	\maketitle
	\begin{abstract}
		\begin{quote}
		Deep generative models for symbolic music are typically designed to model temporal dependencies in music so as 
        to predict the next musical event given previous events. In many cases, such models are expected to learn abstract 
        concepts such as harmony, meter, and rhythm from raw musical data without any additional information. In this study, 
        we investigate the effects of explicitly conditioning deep generative models with musically relevant information. 
        Specifically, we study the effects of four different conditioning inputs on the performance of a recurrent monophonic 
        melody generation model. Several combinations of these conditioning inputs are used to train different model 
        variants which are then evaluated using three objective evaluation paradigms across two genres of music. The results 
        indicate musically relevant conditioning significantly improves learning and performance, 
        and reveal how this information affects learning of musical features related to pitch and rhythm. An informal subjective 
        evaluation suggests a corresponding improvement in the aesthetic quality of generations. 
		\end{quote}
    \end{abstract}
    
    \section{Introduction}\label{intro}
    With recent advances in deep learning, the generation of symbolic music (i.e., music scores) using deep generative models has become quite popular. Common practice is to model music as a sequence and use some kind of Recurrent Neural Network (RNN) to learn the temporal dependencies using a corpus of symbolic music data. 
    
    In its simplest form, this amounts to training a model to predict the next note/event given the previous
    notes/events. Inputs to these models are typically either discrete tokens or one-hot vectors extracted from raw
    symbolic music data with limited or no explicit musical context. For example, monophonic melody generation systems
    typically attempt to learn melody while ignoring any form of explicit harmonic context, which is an important guide
    in note selection. While polyphonic music generation systems operating on piano roll-like formats include this
    information implicitly, the data often lacks harmonic classification and generalization~---~explicit information
    integral to human composers such as inversion,
    % \comment{I don't get this sentence: Why is the information explicit? It
    % is implicit. What is integral to human composers? THat sounds like it's part of the bodies}
     voicing, and harmonic
    function are treated as separate, unconnected events whose abstractions the model is expected to learn on its own.
    Rhythmic considerations like pulse, syncopation, and meter that can play a deciding role in how musical phrases are constructed, are left out. 
    The expectation in these cases is that the deep networks are powerful enough to identify patterns in the data and learn to encode abstract musical concepts such as harmony, meter, and rhythm which is a considerable challenge even for humans. 
    
    In instances where researchers have tried to facilitate learning by explicitly conditioning models with additional 
    music-specific information such as chords \cite{eck2002finding,yang2017midinet} and bar-positions
    \cite{hadjeres2018anticipation,jazzGAN}, these conditionings have been added to the base
    architecture as a means of improving overall performance, rather than tested as independent variables that
    potentially affect particular outcomes. 

    We aim to bridge this gap by explicitly and independently representing various musical components and comparing their effects on the learning and performance of a deep monophonic melody generation system by using them as conditioning inputs. Towards this end, we train and compare the performance of a model consisting of a pair of RNNs which independently model note pitch and note duration sequences conditioned on combinations of musical conditioning inputs. These inputs include
    \begin{inparaenum}[(a)]
    \item inter-conditioning the generation of the pitch and duration networks on each other,
    \item conditioning with the chordal harmony of current and next notes, and 
    \item conditioning with relative bar position.
    \end{inparaenum}
    In total, thirteen different configurations are tested, one for each possible combination of conditioning inputs
    including the case of no conditioning, and evaluated using three different objective
    evaluation metrics using datasets from two different genres of music: 
    \begin{inparaenum}[(a)]
        \item Scottish and Irish folk, and
        \item Bebop Jazz.
    \end{inparaenum}
    
    \section{Related Work}\label{relwork}
    \subsection{Musical Sequence Modeling Using RNNs}
    RNNs have been ubiquitously applied to sequence modeling tasks, and symbolic music generation is no exception \cite{mozer1994neural,franklin2004computational,eck2002finding,colombo2016algorithmic,chu2016song}. Other symbolic music generative models, such as Variational Auto-Encoders (VAE) and Generative Adversarial Networks (GAN), have also included RNNs as integral building blocks \cite{roberts18a,yu2017seqgan}.

    The earliest work on symbolic music generation using neural networks used an architecture designed to take pitch, duration, and chord information as input \cite{mozer1994neural}. The model was inspired by psycho-acoustics, towards which hand-designed embeddings were used for pitches, durations and chords. Most modern deep learning architectures have since adopted trainable embedding layers to learn embeddings for the discrete data symbols based on the raw data. 

    Eck and Schmidhuber were the first to use a Long-Short Term Memory (LSTM) based RNN to generate blues chord progressions with improvised melodies \cite{eck2002finding}. Their experiments generated polyphonic music, and used fixed-length bins to discretize time. 
    %Subsequently Franklin experimented with different representations of pitch and duration and showed that LSTMs were capable of capturing long-term temporal features from music data \cite{franklin2004computational}. 

    %Trieu and Keller's JazzGAN, and Mogren's C-RNN-GAN both train an RNN to predict sequences of
    %notes adversarially, each implementing a different technique to overcome the issue of training a %discrete token generator with outputs from a discriminator \cite{jazzGAN,crnngan}.
    % ASHIS: I don't think the paragraph above adds any value to the paper. 
    
    \subsection{Learning Musical Dependencies}
    Studies in the past have attempted to learn dependencies between musical features by either training and then
    combining separate networks to learn these features independently or by applying them as conditioning inputs.
    
    Chu et al.\ encoded prior musical knowledge in a hierarchically structured RNN~\cite{chu2016song} for generating symbolic scores. They first generate melodies and then use those to generate chords. 
    
    MIDINet~\cite{yang2017midinet} includes explicit chord conditioning, assigning a single triad chord per bar. JazzGAN
    also makes use of explicit chord conditioning, assigning a chord to each half bar in their corpus using 3 distinct
    conditioning methods corresponding to 3 distinct rhythm representations tested in the authors'
    experiments~\cite{jazzGAN}. Notably, in this work we use an identical chord representation to that used by
    JazzGAN (section \ref{datarep}).
    
    Colombo et al.\ proposed an architecture similar to ours, modeling pitch and duration with two separate RNNs \cite{colombo2016algorithmic}. 
    They conditioned their pitch network with the previous pitch and note duration and the duration network with the current pitch and duration. 

    Johnson et al.\ similarly developed a model that integrates separate networks, each of which trains
    on the same musical data represented differently \cite{keller17poe}. Each constituent network outputs a distribution
    over notes/note choices for each timestep; the weighted combination of which is considered as the final output of
    the model. % \comment{rephrase. First of all, shouldn't it be a distribution of notes/note choices, second, what does the 'which' refer to? distributions, networks?}. 
    %In their approach, each constituent network outputs a distribution over the same note choices for a timestep. These outputs are combined using \textit{Product of Experts}\cite{hinton2002poe}. 
    % ASHIS: the above line is confusing and may not be necessary,
    In addition, the authors condition their model using chordal harmony and metric information. 
    % \comment{in case you need to save space, I would remove these references to your stuff that you haven't talked about yet. You can add ashort note below that references the things you note in the related work, but this seems a bit like you start referencing your work before talking about it.}

    Conklin and Witten's  \textit{Multiple Viewpoint Systems for Music Prediction}
    % \comment{this is an awkward start. I assume this is a paper title? Why not just start with the author names}
    derives several sequences of viewpoints, or musical attribute sets, from a sequence of notes in a melody and learns
    a model for each viewpoint independently. The predictions of different viewpoints are combined later into a single
    final prediction \cite{conklin1995views}.
    %These models' predictions of a next note given a set of preceding notes translated into their respective viewpoints are combined into a single final prediction. 
    Cherla et al.\ developed a Restricted Boltzmann Machine
    % \comment{only introduce abbreviations if you use them later} 
    model based on this approach, predicting pitch events using two viewpoints~---~note pitch and note duration \cite{cherla2013distributed}.

    Mogren's C-RNN-GAN learns to jointly predict real-valued tuples of frequency, length, intensity and timing,
    though does not include any chord conditioning \cite{crnngan}.
    % \comment{is it intentional that you use a network name instead of the author names here?}

    MuseGAN, a system for multi-track music generation, also attempts to learn dependencies between different parts of a
    song \cite{dong2018musegan}. The input tracks are represented as piano roll MIDI sequences, and a convolutional GAN based
    architecture is used to learn dependencies between them in three different configurations. 

    % \comment{So the related work is ok, but it still misses a general story. It reads like unconnected paragraphs.}
    \subsection{Data representation for symbolic music}
    There exist several choices for representing symbolic music data. For pitches, the two most popular approaches are:
    \begin{inparaenum}[(a)]
        \item One-hot vectors denoting distinct pitches \cite{colombo2016algorithmic}, or  
        \item Embeddings from a learnable embedding layer that maps pitches to lower dimensional vectors \cite{roberts18a,hadjeres2017deepbach}.
    \end{inparaenum} 
    The latter is inspired by word2vec embeddings \cite{mikolov2013distributed}, which have been used extensively by the
    natural language processing and machine translation communities.

    The most common approach to representing note duration is to split the musical score into fixed-length
    ticks and encode each tick as an event or set of events (e.g. note held, rest, song end) \cite{roberts18a,hadjeres2017deepbach}. 
    While this approach helps maintain a timing-grid and has been used successfully in many studies, it has limitations. 
    For instance, most current models using this approach ignore durations shorter than a sixteenth note. This is because
    representing a finer time resolution supporting shorter notes or uneven time divisions (e.g., triplets) would increase
    sequence lengths significantly, which makes learning with RNNs harder. 

    \section{Method}\label{Method}
    In this study, we aim to analyze the effects of different explicit conditioning inputs on the performance of a recurrent monophonic melody generation system. Specifically, we are interested in how these conditioning inputs impact the prediction of the next note, i.e., predicting the pitch $P_t$ and the duration $D_t$ of the $t^{th}$ note. %given information about the previous notes.

    \subsection{Approach} \label{overall}
    Note pitch and note duration sequences of monophonic melodies are modelled with two parallel LSTM-RNN networks, one for pitch and one for duration. In the experiments, each of these networks is conditioned with various conditioning inputs, the combinations of which define what will hereafter be refered to as conditioning configurations. 
    %Several objective metrics are used to compare and evaluate the performance of these configurations. 
    Before delving deeper into the model architecture and configurations, the data representation used is described. 

    \begin{figure}[t]
        \centering
        \includegraphics[width=3in]{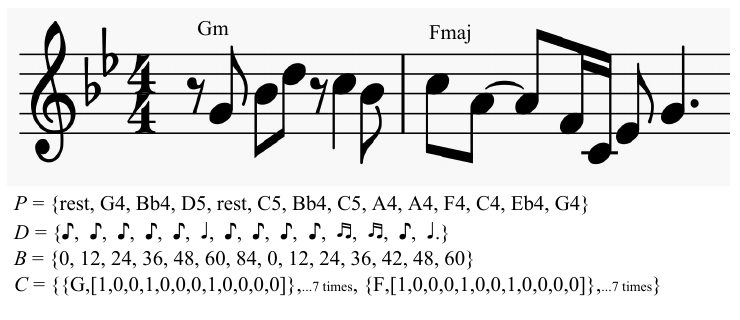}
        \caption{Illustration of data representation scheme. $P$: pitch token sequence, $D$: duration token sequence, $B$: relative bar-position sequence, and $C$: harmony sequence}
        \label{fig:datarep}
    \end{figure}
    \begin{figure*}[t]
        \centering
        \includegraphics[width=0.9\textwidth]{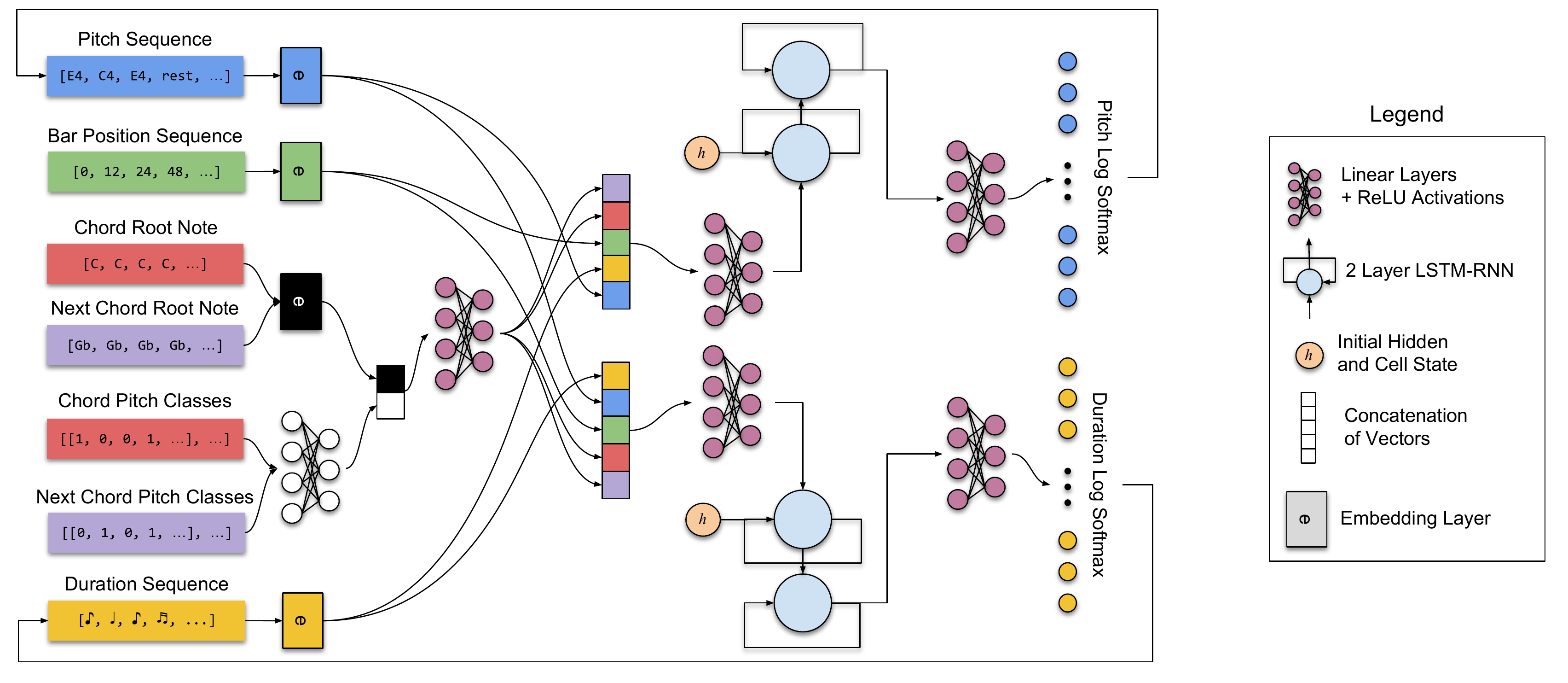}
        \caption{Model architecture overview: This figure displays the \texttt{CNIB} configuration of the model, in which all conditioning inputs are all applied simultaneously. Other configurations can be derived from this by selectively removing a subset of connections prior to the embedding concatenation operation. No information from the current timestep is shared between the  LSTM-RNNs.} 
        %Notably, neither the pitch nor duration network are conditioned using information from the current timestep i.e. the output of one for a timestep is not given to the other to condition the generation of the same step.}
        % ASHIS: Don't think this line is necessary. It is clear from the plot. 
        \label{fig:networkdia}
    \end{figure*}
    
    \subsection{Data Representation} \label{datarep}
    Each monophonic melody $M$ is represented as a sequence of pitch and duration tokens corresponding to itsconstituent notes. Several additional sequences are also constructed from the raw data as described below (see Fig.~\ref{fig:datarep} for an illustration):
    \begin{compactenum}[(a)]
        \item \textbf{Pitch Sequence, $P$}: Pitch tokens comprise of $88$ notes of a standard piano keyboard and a special token for rests. 
        \item \textbf{Duration Sequence, $D$}: Duration tokens are assigned based on a dictionary which maps each of $19$ possible note durations to a duration token. 
        \item \textbf{Harmony Sequence, $C$}: For harmony, the chord type is encoded using a 12-dimensional binary vector with the active pitch classes in the chord. The root pitch is encoded using a separate token.
        \item \textbf{Bar-Position Sequence, $B$}:  Tokens indicating the relative position of each note within a bar. Each beat is divided into 24 equal divisions (this ensures that we can adequately represent triplets) resulting in 96 tokens.
    \end{compactenum}

    \subsection{Model Architecture} \label{model}
    The base model architecture consists of two parallel LSTM-RNN based networks, denoted as
    \begin{inparaenum}[a)]
        \item \textit{Pitch} Network and
        \item \textit{Duration} Network,
    \end{inparaenum}
    which model the pitch sequence $P$ and the duration sequence $D$, respectively. 
    
    The models are trained to maximize the conditional log-likelihood for the current pitch and duration, given information about the previous notes. For the \textit{Pitch} network, this translates to 
    \begin{equation}
        \max_{\theta_P}\sum_{t}logprob(P_t | I_{<t}), t\in[0, L-1]
        \label{eq:1}
    \end{equation}
    where $P_t$ is the pitch of the $t^{th}$ note, $I_{<t}$ is information about all notes that have occurred before the $t^{th}$ note, $L$ is the number of notes, $\theta_P$ represents the parameters of the \textit{Pitch} network.
    Similarly, for the \textit{Duration} network:
    \begin{equation}
        \max_{\theta_D}\sum_{t}logprob(D_t | I_{<t}), t\in[0, L-1]
    \end{equation}
    where $D_t$ is the duration of the $t^{th}$ note, $\theta_D$ represents the paramters of the \textit{Duration} network. All other symbols retain the same meaning as in Equation~\ref{eq:1}.
    
    Information for the $t^{th}$ note is denoted by $I_t$ and is constructed by concatenating embeddings computed for each input in a given conditioning configuration. A list of the conditioning configurations used is provided in Table~\ref{tab:configs}. 
        
    \begin{table}[t]
        \footnotesize
        \centering
        \begin{tabular}{@{}ll@{}}
            \toprule
            \textbf{Conditioning Configuration} & \textbf{Abbreviation} \\ \midrule
            No conditioning & \texttt{No-Cond} \\
            Inter Conditioning & \texttt{I} \\ 
            Chord Conditioning & \texttt{C} \\ 
            Next Chord Conditioning & \texttt{N} \\
            Barpos Conditioning & \texttt{B} \\
            Chord + Inter & \texttt{CI} \\
            Chord + Next Chord & \texttt{CN} \\
            Chord + Barpos & \texttt{CB} \\
            Inter + Barpos & \texttt{IB} \\
            Chord + Next Chord + Inter & \texttt{CNI} \\
            Chord + Next Chord + Barpos & \texttt{CNB} \\
            Chord + Inter + Barpos & \texttt{CIB} \\
            Chord + Next Chord + Inter + Barpos & \texttt{CNIB} \\ \bottomrule
        \end{tabular}
        \caption{List of all conditioning configurations. Each conditioning is implemented by concatenating one/more additional embeddings to the input of the networks.}
        \label{tab:configs}
    \end{table}

    The pitch embeddings $p_t$, duration embeddings $d_t$ and the bar-position embeddings $b_t$ are computed by processing the elements in $P$, $D$ and $B$, respectively, through their respective embedding layers. For chord embeddings $c_t$ and next chord embeddings $n_t$, the root token of the chords are passed through an embedding layer, while the pitch class vectors are encoded using a set of fully connected layers. The outputs of these layers are then concatenated, and subsequently processed together by a second set of fully connected layers to obtain the final chord embeddings. 
    
    In each conditioning configuration, embeddings for the used conditioning inputs are concatenated with the pitch and duration embeddings to form the information vector $I_t$. For example, in the \texttt{No-Cond} configuration of the \textit{Pitch} network, $I_t$ will only contain the pitch embedding, whereas in the \texttt{CI} configuration, $I_t$ will be the concatenation of pitch, current chord, and duration embeddings. 
    
    $I_t$ is then encoded by a set of fully connected layers before being passed through the LSTM-RNN. The output of the LSTM-RNN is decoded by a second set of fully connected layers and activations, followed by a Log-Softmax layer. An overview of the model architecture (\texttt{CNIB} configuration) is shown in Fig.~\ref{fig:networkdia}. All other configurations are a subset of this architecture, i.e, they can be derived by selectively removing certain connections prior to the concatenation operation.

    \section{Experiments}
    A total of $13$ conditioning configurations were used to train models using each of our two datasets of
    chord annotated lead sheets. All models were implemented and trained using PyTorch.
    %\footnote{https://pytorch.org,last accessed: 18th Feb 2019} 
    All code including the training and evaluation scripts is available online.\footnote{https://github.com/bgenchel/Explicitly-Conditioning-Melody-Generation}

    \subsection{Datasets}\label{dataset}
    Training and testing of the models is performed using two monophonic melodic datasets.
    
    \subsubsection{FolkDB}
    FolkDB comprises of chord-annotated lead sheets in the Scottish and Irish folk tradition. The original
    dataset\footnote{https://github.com/IraKorshunova/folk-rnn/tree/master/data, last accessed: 18th Feb 2019} is 
    converted from .abc format to MusicXML.
    %\footnote{https://www.musicxml.com/, last accessed: 17th Feb 2019} 
    Only melodies 
    with $4/4$ time signature and chord annotations were used, resulting in a total of $254$ melodies. 
    
    \subsubsection{BebopDB}  
    For the purposes of this study, we created a new dataset of well annotated Bebop Jazz lead sheets in MusicXML
    format. Each lead sheet contains a monophonic melody with chord annotations, where the chord annotations are placed
    with specific timing relative to a bar.

    At present, the dataset consists of $147$ lead sheets from $61$ composers. The dataset contains $275$ unique chords and $20$ unique chord types (e.g., major7, sus4). The longest song in the dataset contains $96$ bars, while the shortest contains $12$; songs average around $35$ bars. The largest number of notes in a bar is $24$, and the minimum is $1$, with an average of around $6$ notes per bar. % We plan to continue growing this dataset in the future, as well as improving our annotations both in number and quality. 

    \subsubsection{Data Pre-processing} 
    All melodies are transposed to all root notes by shifting each piece up 5 semi-tones and down 6 semi-tones. This results
    in $3042$ songs for FolkDB and $1617$ songs for BebopDB. Transposing the songs, as opposed to shifting each song to the same key, 
    allows the models to better learn songs with key changes. 

    \subsection{Model Specification}
    Embedding dimensions of $8$, $4$, $2$, and $8$ were used for the pitch, duration, chord root, and bar position tokens, respectively. The chord pitch class vector is encoded using linear layers to a dimension of $4$. The concatenation of chord root embedding and chord pitch class encoding is further encoded to a dimension of $8$.
    
    Linear layers appear in pairs, with the exception of single linear layers feeding into the LSTM-RNNs. 
    The output size of the first layer is the arithmetic mean of its input and the output of the second layer.
    %The first layers project their input to a size equal to the arithmetic mean of their input size and the output size of the second layers. 
    The layers are separated by a batch-normalization layer \cite{ioffe2015batchnorm} followed by a ReLU activation. Batch-Norm and ReLU layers also follow the single linear layer preceding the LSTM.

    Each model uses uni-directional, 2-layer LSTM-RNNs for which both the input size and hidden size are $256$-dimensional. The output of the \textit{Pitch} LSTM-RNN is decoded to an 89-dimensional vector, while the \textit{Duration} LSTM-RNN output is decoded to a 19-dimensional vector. Each decoding is fed to a Log-Softmax layer.

    \subsection{Training Specification}
    Each model is trained by feeding in sub-sequences of $64$ notes ($I_{n}:I_{n+64}$) and backpropagating the Negative Log-Likelihood Loss (NLL) taken between their outputs and the following sub-sequence of $64$ notes from the ground truth data ($I_{n+1}:I_{n+65}$). For training, overlapping windows of length $64$ are extracted from each song in the datasets with a hop-size of $1$. 
    
    Both \textit{Pitch} and \textit{Duration} networks are trained for $30$ epochs using a batch-size of $64$. We found 30 epochs sufficient for each configuration to converge without overfitting, and thus, used it as an early stopping point. Datasets are divided into training and validation sets using an 80/20\% split. We use the AMSGrad variant \cite{reddi2018convergence} of the Adam optimizer \cite{kingma2014adam} with a learning rate of $1e-3$  
    %$\beta_1 = 0.9$, $\beta_2 = 0.999$ and $\epsilon = 1\mathrm{E}{-8}$. 
    Models are regularized by applying dropout to the LSTM layers with $0.2$ probability. 

    \section{Evaluation}\label{evaluation}
    The performance of the configurations is evaluated using three separate objective metrics: 
    \begin{inparaenum}[(a)]
        \item NLL loss achieved on the validation set,
        \item performance on the MGEval framework \cite{yang_evaluation_2018}, and
        \item BLEU score \cite{papineni2002bleu},
    \end{inparaenum}
    followed by an informal subjective appraisal of melodies generated by each configuration.

    The first metric, NLL, is the most common metric to measure the predictive capability of a generative model and its
    overall efficiency. The other two, though used to a lesser extent, measure the degree to which the generated melodies comply with the statistics of the training data.

    \subsection{NLL Loss}
    The conditioning configurations are trained to predict the pitch and duration of the current note given information about previous notes. For comparison, we look at the lowest NLL loss obtained on the validation set by each configuration. Table~\ref{tab:nll} summarizes the results for all different configurations. 

    \begin{table}[t]
        \footnotesize
        \centering
        \begin{tabular}{@{}lcccc@{}}
        \toprule
        Validation NLL & \multicolumn{2}{c}{\textbf{FolkDB}} & \multicolumn{2}{c}{\textbf{BebopDB}} \\ \midrule
        \textbf{Configuration} & \textit{Pitch} & \textit{Duration} & \textit{Pitch} & \textit{Duration} \\ \toprule
        \texttt{No-Cond} & 0.373 & 0.143 & 0.381 & 0.136 \\
        \texttt{I}       & 0.303 & 0.061 & 0.275 & 0.065 \\
        \texttt{C}       & 0.317 & 0.094 & 0.313 & 0.085 \\
        \texttt{N}       & 0.319 & 0.083 & 0.318 & 0.075 \\
        \texttt{B}       & 0.277 & 0.117 & 0.276 & 0.112 \\
        \texttt{CI}      & 0.264 & 0.055 & 0.241 & 0.056 \\
        \texttt{CN}      & 0.313 & 0.075 & 0.304 & 0.074 \\
        \texttt{CB}      & 0.240 & 0.083 & 0.242 & 0.073 \\
        \texttt{IB}      & 0.307 & 0.053 & 0.274 & 0.057 \\
        \texttt{CNI}     & 0.248 & \textbf{0.052} & \textbf{0.226} & \textbf{0.050} \\
        \texttt{CNB}     & \textbf{0.212} & 0.067 & 0.239 & 0.065 \\
        \texttt{CIB}     & \textbf{0.217} & \textbf{0.048} & \textbf{0.198} & \textbf{0.048} \\
        \texttt{CNIB}    & \textbf{0.206} & \textbf{0.047} & \textbf{0.190} & \textbf{0.045} \\ \bottomrule
        \end{tabular}
        \caption{Best Validation NLL (lower is better) results during training for the \textit{Pitch} and
        \textit{Duration} networks of different conditioning configurations. Adding different conditioning inputs gradually improves NLL. Adding barposition (\texttt{B}) works better for \textit{Pitch} networks, inter-conditioning (\texttt{I}) works better for \textit{Duration} networks. {Bold} items denote the top $3$.}
        \label{tab:nll}
    \end{table}
    \begin{figure}[b!]
        \centering
        \includegraphics[width=3in]{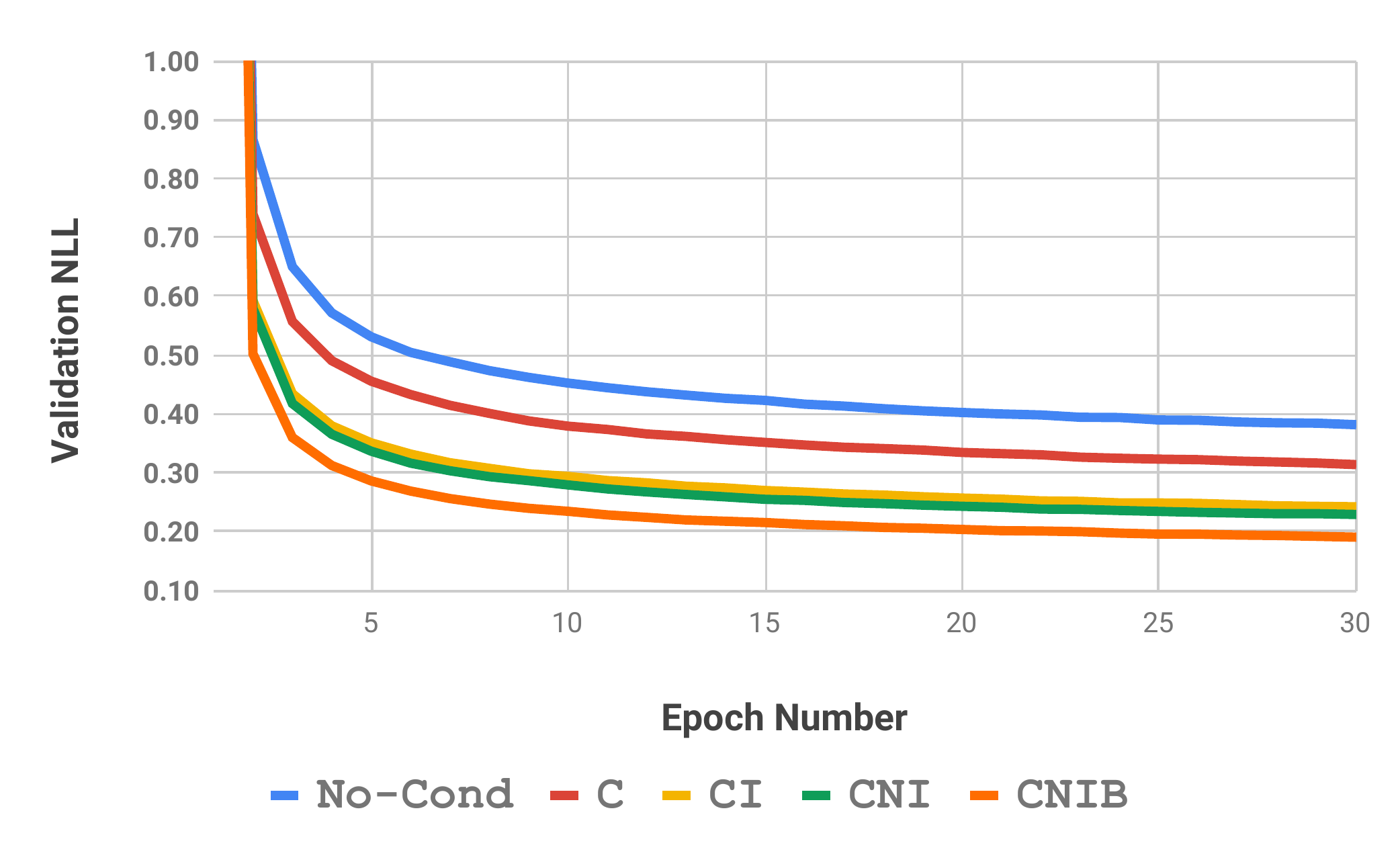}
        \caption{Validation NLL curves for $5$ different configurations for the \textit{Pitch} network trained on BebopDB. Adding conditioning inputs improves speed (faster convergence) and performance (lower value) during training.}
        \label{fig:losscurves}
    \end{figure}

    \subsubsection{Discussion}
    The results suggest that the addition of greater numbers of conditioning inputs improves the predictive performance of both \textit{Pitch} and \textit{Duration} networks on both datasets. 
    This is further supported by a comparison of validation loss curves across training epochs as shown in
    Fig.~\ref{fig:losscurves}, where it is clear that more conditioned models converge to lower values faster. This is
    expected, as the network is provided with more musically relevant context which dictates the flow of the melody. 
    Some observations are:
    \begin{compactenum}[(a)]
        \item Adding bar position \texttt{B} significantly improves the performance of \textit{Pitch} networks on this metric for
            both datasets. This indicates that explicit encoding of this information facilitates better learning
            relative to implicitly specifying this information by providing only note durations as input. It is
            interesting to note that bar position does not improve the performance of \textit{Duration} networks when
            compared to the effects of other inputs such as previous pitch (inter) \texttt{I}, chords \texttt{C}, or next chords \texttt{N}.
        \item Inter-conditioning the \textit{Duration} network improves its performance significantly across both datasets on this metric. 
            This indicates that predicting the duration of the current note relies more heavily on the previous pitch than the current chord, next chord, or bar position. 
        \item Providing information regarding the current and next chords without other conditioning inputs (\texttt{CN}
            case) does not appear to improve performance on this metric, indicating that information regarding the previous pitches/durations and bar positions are possibly more important for learning and prediction.
    \end{compactenum} 

    \subsection{MGEval}
    The MGEval (Music Generation Evaluation) toolbox was designed by \citeauthor{yang_evaluation_2018} for objective
    evaluation of music generation systems \cite{yang_evaluation_2018}. It uses several pitch and duration-based
    features to measure the degree to which the generated music is able to match the statistics of the training data.
    The features are modeled as probability distributions. The performance of a generative model is
    evaluated by computing the distance between these probability distributions across the training data and generated
    melodies using KL-divergence. 
    
    This toolbox is used to compare the performance of each conditioning configuration and to
    further analyze the impact of each condition (\texttt{I}, \texttt{C}, \texttt{N}, and \texttt{B}).
    We first generate new melodies using our conditioning configurations for each non-transposed song in both
    datasets. Melodies are generated by feeding an initial seed of 10 notes to the trained models
    and then recurrently sampling pitch/duration values from the output Softmax distributions. For models requiring
    chords, we use the chord progressions from the lead sheets. A set of $11$ features, sub-categorized into pitch and
    duration types, is computed (see Table~\ref{tab:mgeval_feats}) for both the generated
    melodies and the original melodies in the datasets. The KL-Divergence between the predicted distribution and
    inter-set distribution is used as the performance metric.

    \begin{table}[t]
        \footnotesize
        \centering
        \begin{tabular}{p{0.5in}p{2.5in}}
            \toprule
            \textbf{Feature Type} & \textbf{Feature Name} \\ \midrule
            Pitch-based    & Pitch Count (PC), Pitch Count/Bar (PC/bar), Pitch Class Histogram (PCH), Pitch Class Transition Matrix (PCTM), Pitch Range (PR), average Pitch Interval (PI) \\ \midrule
            Duration-based & average Inter-Onset Interval (IOI),  Note Length Histogram (NLH), Note Length Transition Matrix (NLTM), Note Count (NC), Note Count/Bar (NC/Bar)        
            \\ \bottomrule
        \end{tabular}
        \caption{List of features used from the MGEval Toolbox. More information about these features and their computation process is given in the original paper \cite{yang_evaluation_2018}.}
        \label{tab:mgeval_feats}
    \end{table}

\begin{figure}[t!]
        \centering
        \includegraphics[width=2.5in]{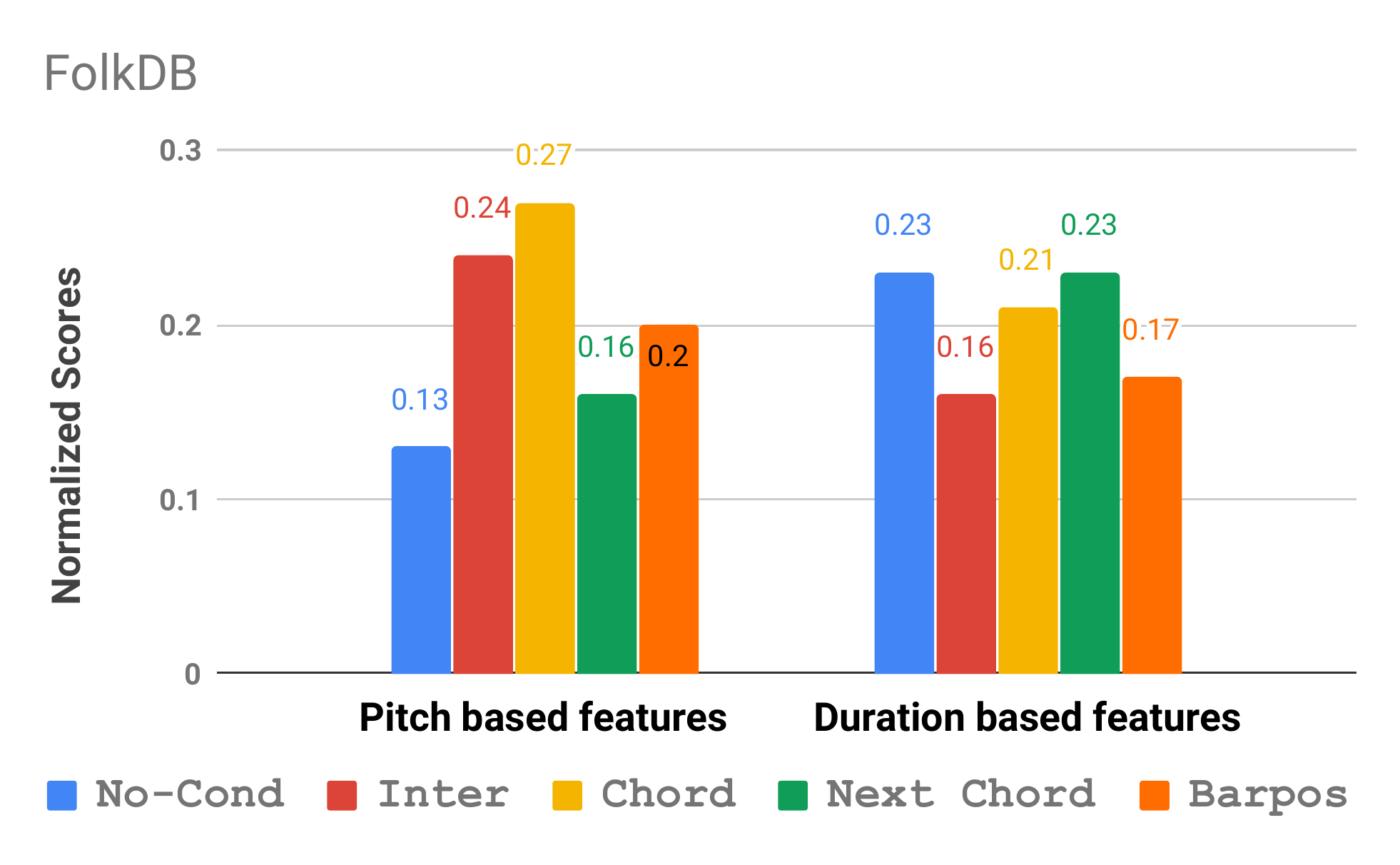}
        \includegraphics[width=2.5in]{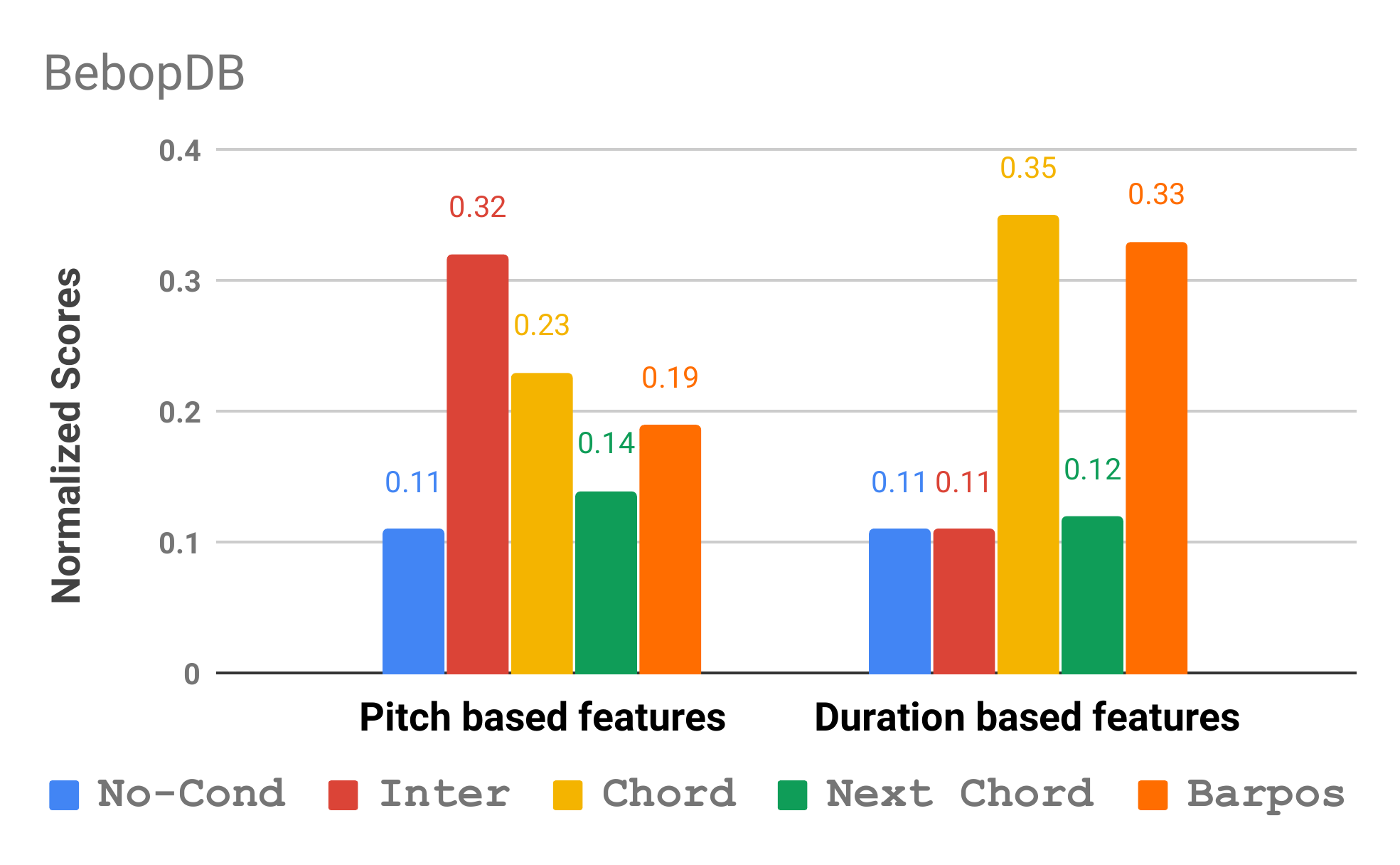}
        \caption{Aggregated score distribution (higher is better) across MGEval sub-categories for each conditioning input} 
        \label{fig:mgeval-agg}
    \end{figure}

    Table~\ref{tab:mgeval_results} shows the performance of two such features. Since there are a large number of features and conditioning configurations, we use an aggregation technique to summarize the complete set of results.\footnote{https://bgenchel.github.io/ecmg/results} For each feature, the best performing configuration (lowest KL-divergence) is given a score of $1$. If there are several configurations with very close KL-divergences (within one standard deviation of the distribution), the score is equally split. Next, the scores are allotted to the individual conditioning inputs based on the conditioning configuration. For instance, for the \texttt{CNI} configuration, the score is split equally amongst \texttt{C}, \texttt{N} and \texttt{I}. The scores for each conditioning input are then added across features within each feature sub-category and normalized to a sum of one. The final normalized scores can be interpreted as the probability that a particular conditioning performs better than others. These normalized distributions are presented in Fig.~\ref{fig:mgeval-agg}. 

    \begin{table}[t]
        \footnotesize
        \centering
        \begin{tabular}{@{}lcccc@{}}
        \toprule
        KL-Divergence & \multicolumn{2}{c}{\textbf{FolkDB}} & \multicolumn{2}{c}{\textbf{BebopDB}} \\ \midrule
        \textbf{Configuration} & \textbf{PR} & \textbf{NLH} & \textbf{PR} & \textbf{NLH} \\ \toprule
        \texttt{No-Cond} & \textbf{0.019} & \textbf{1.652} & \textbf{0.027} & \textbf{2.154} \\
        \texttt{I}       & \textbf{0.035} & 2.297 & \textbf{0.029} & 4.750 \\
        \texttt{C}       & 0.046 & 2.212 & 0.047 & \textbf{2.452} \\
        \texttt{N}       & 0.050 & 2.696 & 0.048 & 3.717 \\
        \texttt{B}       & 0.052 & 2.799 & 0.060 & \textbf{2.731} \\
        \texttt{CI}      & 0.094 & 2.671 & 0.046 & 5.347 \\
        \texttt{CN}      & 0.059 & 3.063 & 0.061 & 3.588 \\
        \texttt{CB}      & 0.072 & 2.273 & \textbf{0.023} & \textbf{2.409} \\
        \texttt{IB}      & 0.073 & 2.337 & \textbf{0.022} & 4.471 \\
        \texttt{CNI}     & 0.090 & 3.368 & 0.038 & 5.584 \\
        \texttt{CNB}     & 0.057 & 2.792 & 0.050 & \textbf{2.914} \\
        \texttt{CIB}     & \textbf{0.033} & \textbf{1.439} & \textbf{0.030} & 6.101 \\
        \texttt{CNIB}    & 0.056 & \textbf{1.883} & 0.041 & 5.965 \\ \bottomrule
        \end{tabular}
        \caption{Predicted-set to inter-set KL-divergence (lower is better) for two MGEval features.
        PR: Pitch Range, NLH: Note Length Histogram. {Bold} items are best performers for that feature (within
        one standard deviation of the top performer).}
        \label{tab:mgeval_results}
    \end{table}

    \subsubsection{Discussions}
    On an average, conditioning inputs improve the performance on the MGEval features. However, different conditioning inputs affect the different feature sub-categories differently. Inter, chord and bar-position clearly perform better for pitch-based features. This is in line with our observations for the NLL loss. There are, however, no clear winners for the duration based features. For FolkDB, next-chord performs the best, while for BebopDB, chord and bar position perform better. Interestingly, for FolkDB, the no-cond case scores the highest for duration-based features. This might be due to the simple rhythmic patterns in that genre.

    While these results are interesting, there are also a few concerns. Most prominently, it is assumed that there is an equal contribution from all inputs towards the performance of configuration. To further verify, all scores were normalized for all metrics to have zero mean and unit varience, and t-tests were conducted for each condition for both pitch-based and rhythm-based features. These tests showed that the results are statistically significant for inter, chord, and next chord conditioning in BebopDB and for chord and next chord conditioning in
    FolkDB. It is also interesting to note that Table~\ref{tab:mgeval_results} does not display the same trend as Table~\ref{tab:nll}, in which improvement was clearly correlated with increased conditioning. Additionally, the relatively weak performance of inter-conditioning for duration-based features runs counter to the observation of the NLL losses. This indicates that while model training improves with different conditioning inputs, the musical properties of the generated melodies with respect to the training data do not necessarily improve.

    \subsection{BLEU Score}
    The BLEU score, a measure of similarity between two corpuses of text, was originally designed to give an objective evaluation for machine translation tasks \cite{papineni2002bleu}. It is calculated as the geometric mean of the counts of matching n-grams between the generated and target corpuses. A perfect score of $1$ is attained only when the two corpuses are exactly the same, which is almost impossible even for humans to achieve on a generative task. Even though the utility of this metric for generative tasks is debatable, BLEU score can still be considered a useful objective metric in the absence of better objective metrics to quantify creativity and musicality. Therefore, it has been used to evaluate music generation systems \cite{yu2017seqgan}. 
    
    We compute corpus-level BLEU scores (shown in Table~\ref{tab:bleu}) using $1$, $2$, $3$ and $4$-gram sequences for the generated 
    melodies for each configuration. The melody generation process is the same as that followed for the MGEval case. 

    \begin{table}[t]
        \footnotesize
        \centering
        \begin{tabular}{@{}lcccc@{}}
        \toprule
        BLEU Score & \multicolumn{2}{c}{\textbf{FolkDB}} & \multicolumn{2}{c}{\textbf{BebopDB}} \\ \midrule
        \textbf{Configuration} & \textit{Pitch} & \textit{Duration} & \textit{Pitch} & \textit{Duration} \\ \toprule
        \texttt{No-Cond} & 0.267 & \textbf{0.874} & 0.098 & \textbf{0.875} \\
        \texttt{I}       & 0.269 & 0.717 & 0.101 & 0.568 \\
        \texttt{C}       & 0.297 & 0.782 & 0.133 & \textbf{0.688} \\
        \texttt{N}       & 0.311 & \textbf{0.796} & 0.112 & 0.663 \\
        \texttt{B}       & 0.242 & 0.763 & 0.112 & \textbf{0.696} \\
        \texttt{CI}      & 0.278 & 0.685 & 0.130 & 0.569 \\
        \texttt{CN}      & \textbf{0.326} & 0.779 & \textbf{0.153} & 0.666 \\
        \texttt{CB}      & 0.289 & 0.784 & 0.121 & 0.654 \\
        \texttt{IB}      & 0.253 & 0.689 & 0.129 & 0.607 \\
        \texttt{CNI}     & \textbf{0.326} & 0.684 & \textbf{0.157} & 0.572 \\
        \texttt{CNB}     & \textbf{0.335} & \textbf{0.794} & \textbf{0.146} & 0.638 \\
        \texttt{CIB}     & 0.283 & 0.694 & 0.132 & 0.593 \\
        \texttt{CNIB}    & \textbf{0.317} & 0.688 & 0.128 & 0.566 \\ \bottomrule
        \end{tabular}
        \caption{Corpus Level BLEU Scores (higher is better) for \textit{Pitch} and \textit{Duration} networks for different conditioning configurations and datasets. \textbf{Bold} items denote the top $3$.}
        \label{tab:bleu}
    \end{table}

    \subsubsection{Discussion}
    Between the \textit{Pitch} and the \textit{Duration} networks, the scores are more intuitively understandable for the former case. For \textit{Pitch} networks, the highest scores are achieved by configurations which include chords, either current or next. This indicates that adding chords as conditioning inputs helps the network to better reproduce pitch n-grams. 
    
    For the \textit{Duration} networks, \texttt{No-Cond} performs the best. This is quite surprising since this configuration only relies on the duration of previous notes. The second and third highest scores come from chord and bar position conditioning. The FolkDB models work better with next chord, while the BebopDB models perform better with the current chord. Interestingly, once again, the lowest scores for duration seem to correspond with inter-conditioning. 
    
    The results, especially those for \textit{Duration} networks, run counter to the NLL loss results reported above. However, this tallies with observations from the MGEVal framework and warrants further investigation.

    \subsection{Subjective Appraisal of Sample Melodies}
    In order to get a sense of how well the objective measures correspond to our aesthetic experience, we generate a few melodies following the harmony and seed selected randomly from a few lead sheets. While there were some similarities between generations of FolkDB and BebopDB models based on the conditioning configurations, due to musical differences between the genres, some were harder to judge than others.
    
    Models which lack any kind of chord conditioning were in general able to stay in key for both genres for some time. As Folk songs tend to stay within the same key throughout, chord conditioning did not seem necessary for consonance. For Bebop generations however, configurations that lacked chord conditioning, eventually fell out of key and were unable to find their way back. 

    For Folk generations, the addition of inter-conditioning seemed to induce specific bad habits, such as repeatedly returning to a particular note, or making repeated large jumps to note B4. These were mitigated in combination with chord conditioning, which in general led to higher quality generations. For Bebop generations, we saw no such effect for inter-conditioning in isolation, however, we did observe this synergy when combined with chord conditioning.

    Barpos conditioning appeared to enhance rhythmic variety in Bebop generations and increase the duration for which generations stayed in key. The same was not observed for Folk generations. Chord conditioning produced the most noticeable effect for both genres, significantly increasing harmonicity for Bebop generations and improving the musicality of Folk generations. Chord-conditioned Folk generations seemed to include a greater number of intervals such as major and minor thirds and well-placed fourths, which we felt added more emotion and musicality. Without chord conditioning, Folk generations tended to over-use consecutive whole tone (major 2nd) intervals which sounded monotonous. In addition, chord conditioning seemed to have a strong positive impact on rhythmic phrasing and variety for both genres. 

    Chord conditioning seemed to have a much more positive impact on Bebop generations than next chord conditioning, which sometimes appeared to cause the model to step out of key, perhaps in anticipation of the next chord. For Folk generations, chord and next chord conditioning appeared to similarly improve performance. 

\begin{figure}[!h]
        \centering
        \includegraphics[width=3in]{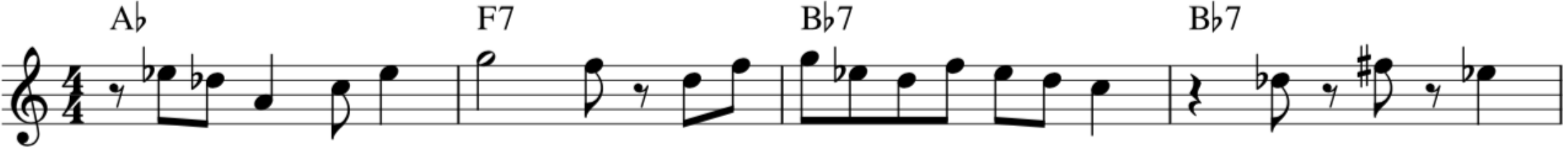}
        \includegraphics[width=3in]{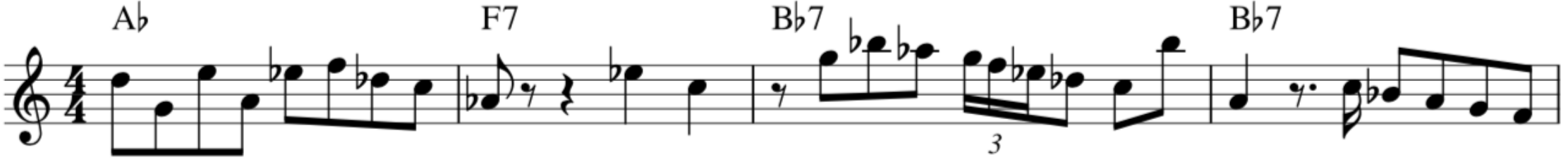}
        \includegraphics[width=3in]{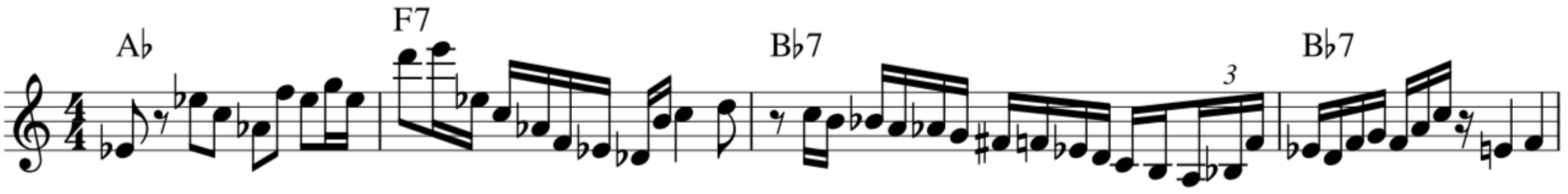}
        \caption{Generations over the first 4 bars of ''Donna Lee'' by Charlie Parker for (from top to bottom) \texttt{I}, \texttt{CI}, and \texttt{CIB}.}
        \label{fig:score-samples}
    \end{figure}

    Despite the observed effects listed above, we still found that \texttt{CNIB} and \texttt{CIB} produced the most aesthetically pleasing melodies over all, indicating positive synergy between conditionings even when they appear detrimental in isolation. \texttt{CI}, \texttt{CNI}, and \texttt{CNB} were close seconds in quality. These observations are consistent with results from the objective metrics, indicating the importance of harmony on both pitch and rhythmicity, as well as its synergy with inter-conditioning and to a lesser extent bar position. Selected score samples are displayed in Figure \ref{fig:score-samples}. We provide audio for these generated melodies for the reader to make their own judgments\footnote{https://bgenchel.github.io/ecmg/results}.
% \comment{Would it be worthwhile to link the objective results to your subjective results, maybe give a metric preference?}

    \section{Conclusion}\label{conclusion}
    We present a case study designed to evaluate and analyze the effects of explicit musical conditioning on monophonic melody generation. We describe thirteen different conditioning configurations of four musical conditioning inputs (inter-conditioning between pitch and duration sequences, current chord, next chord, and bar position), each of which is used to train an LSTM-RNN based model on two datasets of chord-labeled melodies. We evaluate the performance of models in each conditioning configuration with three objective measures, providing insight into how these conditioning inputs effect overall learning rate and accuracy and how they facilitate the model in learning musical features from the data. In addition, we also provide a subjective appraisal of the aesthetic qualitty of the melodies generated by models trained using different conditioning configurations.

    The results of this study suggest that harmonic conditioning is important not just for pitch prediction, but for duration prediction as well. More generally, it provides some insight into the relative usefulness of each of the conditioning inputs for learning pitch and rhythm-based features. We also discover that while some features may appear ineffective or even detrimental when applied individually, or in certain combinations, they may still be useful when applied in other combinations. In addition, we show that the effectiveness of musical conditioning is also dependent on the type of data used; though there were commonalities in how conditioning effected generations for both Folk-trained and Bebop-trained models, there were significant differences as well.

    In the future, we aim to conduct a subjective listening test to formally determine if the results of our objective metrics line up with the aesthetic perception, as well as test a greater number of musical conditioning factors, some of which relate to long term structure.

    \bibliographystyle{mume}
    \bibliography{mume2019}
\end{document}